# Distributed Simulation Platform for Autonomous Driving

Jie Tang[1], Shaoshan Liu[2], Chao Wang[3], Quan Wang[3],

[1] South China University of Technology, China
[2] PerceptIn
[3] Baidu USA
{cstangjie@scut.edu.cn, shaoshan.liu@perceptin.io, wangchao30@baidu.com, wangquan02@baidu.com}

**Abstract.** Autonomous vehicle safety and reliability are the paramount requirements when developing autonomous vehicles. These requirements are guaranteed by massive functional and performance tests. Conducting these tests on real vehicles is extremely expensive and time consuming, and thus it is imperative to develop a simulation platform to perform these tasks. For simulation, we can utilize the Robot Operating System (ROS) for data playback to test newly developed algorithms. However, due to the massive amount of simulation data, performing simulation on single machines is not practical. Hence, a high-performance distributed simulation platform is a critical piece in autonomous driving development. In this paper we present our experiences of building a production distributed autonomous driving simulation platform. This platform is built upon Spark distributed framework, for distributed computing management, and ROS, for data playback simulations.

## 1 Introduction

Autonomous driving systems usually consist of sensing, perception, decision making, control and other functional modules, and each module has its own intricate structures and algorithms [1]. In most cases, it is difficult for system or algorithm developers in the testing process to evaluate the massive design space. To test any algorithm change, developers need to test a functional module alone, and later on setting up a whole physical testing environment that consists of a number of other modules, leading to enormous testing costs. Fortunately, many of the testing tasks can be accomplished by utilizing simulator. The key to the success of a simulation is how accurately the simulator can simulate the physical reality.

There are two main kinds of simulation technologies: the first one simulates the environment based on synthetic data, this kind of simulators are mainly used for control and planning, especially at the initial development stage of the algorithms. The second type is based on real data playback to test the function and performance of the different components, which is used mainly in the iterative process of algorithm development. In this article, we mainly discuss the simulators based on data playback.

In order to simulate the environment as realistic as possible, our simulator is built upon the Robot Operating System (ROS), which is used in physical autonomous driving systems [2, 17]. ROS is a distributed computing framework based on message delivery, which makes it easier for developers to make modular programming. Its modular design is critical for the design of simulators since we usually test modules independently. In autonomous driving systems, each functional module in the ROS is deployed in a node, and the communication between the nodes relies on the messages with well-defined formats, *e.g.* messages that contain images. Therefore, developers only need to use the same communication format, and develop simulation module for each functional module, and finally match real functional modules and the simulated modules based on test requirements. For example, if we want to coordinate the functions of the decision module and the control module, we need to install the decision module, control module and other simulated modules into the simulator for testing. If the decision-making module needs to test the new decision-making algorithm separately, we can only install the latest decision module with the other simulated modules on the simulator. The result of this test is only for the decision-making module.

### 1.1 Anatomy of Autonomous Driving Simulators

Firstly, the autonomous vehicle simulator contains a dynamic model of the car, which is used to load the test of autonomous driving system and simulates the behavior of the autonomous vehicle itself. Secondly, the simulation of the external environment is needed, which includes static and dynamic scenes. Static scenes include a variety of stationary traffic signs, such as stop lines, traffic signs, etc. Dynamic scenes mainly refer to the dynamic traffic flow models around the car, such as vehicles, pedestrians, traffic lights and so on. All of these elements construct an analog world corresponds to the real world.

### 1.2 Applications of Autonomous Driving Simulators

In the real world, autonomous vehicles face complex and varied external environments. A good simulator decomposes external environment into the basic elements, and then rearranges the combination to generate a variety of test cases, each simulating a specific scenario. Take a simple set of test cases. Figure 1 shows a simple simulation scene, in which we need to test the response of an autonomous vehicle to a car in front of it, or the barrier car. The initial position of the barrier car is a simulation variable, such that in this case, it may appear from the left front, left, left rear, front, rear, right front, right, right rear relative to the autonomous vehicle, eight directions in total. Next, the speed of the the barrier car is another simulation variable, which can be divided into three categories, faster than the autonomous vehicle, equal to the speed of the autonomous vehicle, and slower than the autonomous vehicle. The next motion step of the barrier car is yet another simulation variable, which can be divided into going straight, turning to the left, and turning to the right. By multiplying all these simulation variables and removing all the unwanted cases, we get a set of test cases.

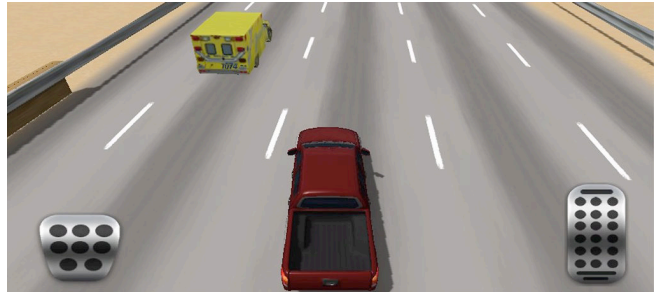

**Fig 1. A simulation scene**

### 1.3    Challenges of Autonomous Driving Simulators

The core problem of the simulator lies in how realistic we can simulate the actual driving environment.   No matter how good the simulator is, the artificial simulation of the scene and the real scene still have some differences. There are still many unexpected events in the real scene that cannot be simulated in a simulator. Therefore, if you can use the real traffic data to reproduce the real scene, you will get better test results compared to the artificial simulation of the scene.  However, the major problem of replaying real-world data is the computing power required to process the massive amount of real world data. If we want to reproduce the scene of every section of the real world on the simulator, we need to let the autonomous vehicles collect the information of each section of the road.  This amount of information cannot be processed on single machines.  Furthermore, in each scene, we can further break it down into basic fragments and to rearrange the combinations of these fragments to generate more test cases.   However, this would only generate even more data and add more burden to the already stressed simulation platform. In this paper, we present the first generic distributed simulation platform for autonomous driving simulation.

## 2    A ROS-based Autonomous Driving Simulator

ROS is a robot operating system based on messaging communication. Its communication mode can be abstracted as a message pool architecture, the message sending node transfers the advertise method to send ROS message to the specified Topic, and the message receiving node transfers the subscribe method to receive the ROS message from the specified Topic.

## 2.1 ROSBAG

Rosbag is a tool that uses this architecture to record from Topic and replay the ROS message to Topic, which is used in the data collection process for unmanned vehicle. Its function is divided into two categories: Record and Play. The Record function is to create a recording node in the ROS, and call the subscribe method to receive ROS message to all the Topics or the specified ones, and then write the message to the Bag file. While the Play function is to establish a play node in ROS, and call the advertise method to send the message in bag to the specified Topic according to timeline.

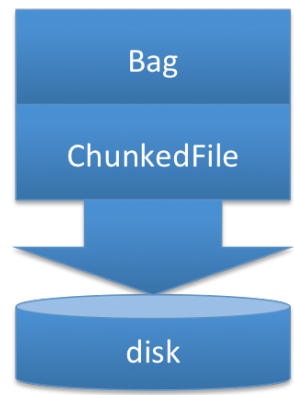

**Fig 2. ROSBag design**

The data format that produced by Rosbag is Bag, which is a file format with two-tier logical structure. As shown in Figure 2, the upper class of the Bag class provides a method for user to operate the file on the abstraction, the down class packages operation methods to the ChunkedFile. ChunkedFile class mainly stores the data separately, and the stored data is a section The latter mainly contains of images or 3D point cloud scan file data collected by autonomous vehicle sensors. Therefore, with ROS, we can easily process, understand and persist multimedia data. However, this presents a challenge to the distributed computing framework, which by default only processes text-based data. We will discuss this issue in detail in the next section.

## 2.2 Simulation Dataset

As we have mentioned before, we mainly focus on the simulators based on real data playback. The first question is the scale of the real-world data. To understand this we can start with the KITTI dataset [12]. In this dataset, KITTI researchers recorded real data for 6 hours with a data volume of 720GB. However, the 6-hour of data is only enough to perform some simple verification tests on algorithms, and it is far from enough to perform full production simulations. To perform full scale production simulation, for example, Google's autonomous driving project collects more than 40,000 hours of real data in the past few years, the total amount of data is estimated to exceed 5 PB. Performing simulations on single machines can not handle data at such scale, and therefore we must design an efficient distributed computing platform based on the real data playback simulators

### 2.3 The Demand on Computing Power

The huge amount of data processing imposes enormous pressure on the computing platform. For instance, the original data for the KITTI data set for 6 hours includes more than 100 million 140-megapixel color charts, and we use a single-machine simulation system to perform deep-learning based segmentation tasks, processing each image takes about 0.3 seconds. In this way, it takes more than 100 hours to analyze the KITTI dataset alone, and if we analyze the whole image dataset for Google's autonomous driving project for example, it will take more than 600,000 hours on to process one full round of simulation on a single machine.

## 3 A Spark-based Distributed Simulation Platform

We have decided to use distributed computing to process simulation in parallel, and we choose Spark as our distributed computing platform. Spark is a universal parallel computing framework opening source by UC Berkeley AMPLab [3]. Spark's distributed computing is based on RAM, which provides significant performance advantages over Hadoop, which persists intermediate data on disks [13]. Unlike Hadoop, the Spark Job's intermediate output and results can be stored in memory, so there is no need to read and write HDFS [14], as a result Spark can be better applied to Map-Reduce algorithm which requires intensive iterative computing.

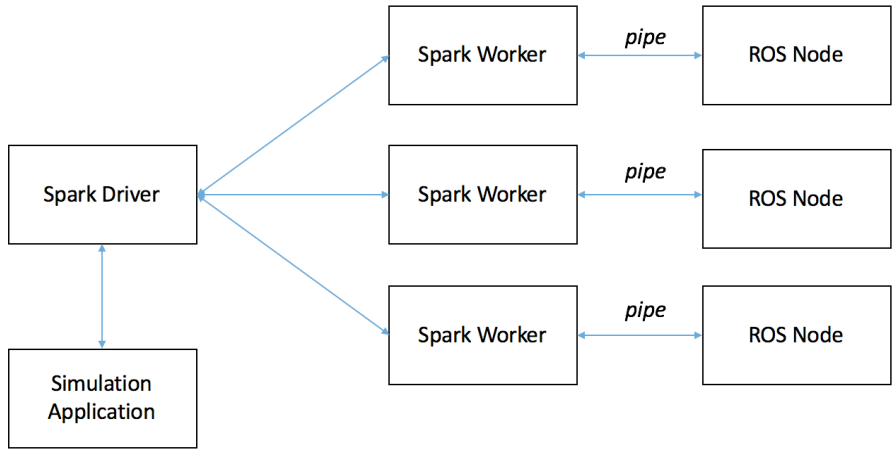

**Fig 3. Architecture of distributed simulation platform**

As Figure 3 has shown, we design and implement a distributed simulation platform framework which is based on Spark to perform autonomous vehicle playback simulation efficiently. We use Spark to manage resource allocation, data input output, and management of ROS nodes. On the Spark driver, we can launch different simulation applications, such as localization algorithms that consume LiDAR raw data, object recognition algorithms that consume image data, vehicle decision-making and control algorithms *etc*. The Spark Driver allocates resource from the Spark worker based on the requested amount of data and computation. Each Spark worker first reads the Rosbag data into memory and then launches a ROS node process the incoming data.

The interface between Spark and ROS is one design decision we need to make. The first approach is to use JNI [15] to connect Spark worker and ROS Node, but this involves the modification of ROS, making the whole system difficult to maintain and evolve. The second approach is to use Linux pipes [16], which create a unidirectional data channel that can be used for inter-process communication. Data written to the write end of the pipe is buffered by the kernel until it is read from the read end of the pipe. We choose to use the second approach since this is easier to maintain. In the design of the pipe, there are two issues that need to be solved: first, Spark only supports consuming text-based data by default, and it does not support multimedia data consumption. We need to design an efficient method for it to consume binary file. Second, we need a way to read from the memory of the cache data through ROSBag play function, and also a way to cache data into memory through ROSBag record function.

### 3.1 Binary Data Streaming

The core of Spark's data structure is Resilient Distributed Datasets (RDD), which allows programmers to perform memory calculations on a large cluster in a fault-tolerant manner. To solve the problem of having Spark consuming multimedia data, we develop a new RDD, the BinPipedRDD, which is shown in Figure 4 below.

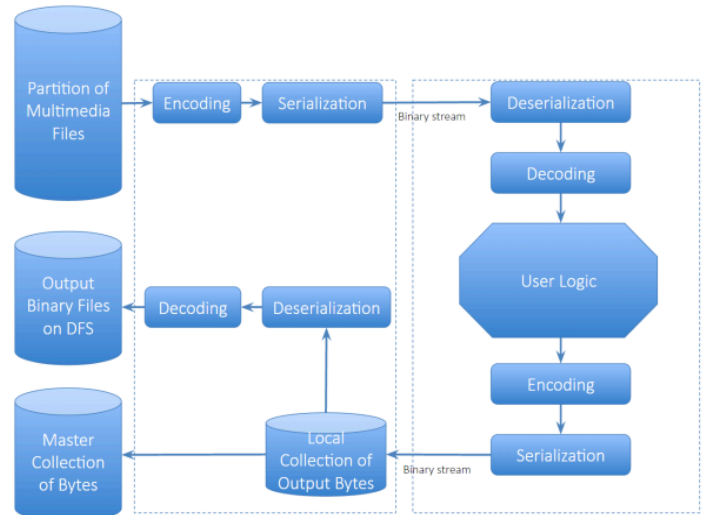

**Fig 4. BinPiped RDD Design**

First, the partitions of binary files go through encoding and serialization stages to form a binary byte stream. The encoding stage will encode all supported inputs format including strings (e.g., file name) and integers (e.g., binary content size) into our uniform format, which is based on byte array. Afterward, the serialization stage will combine all bytes arrays (each may correspond to one input binary file) into one single binary stream. Then, the user program, upon receiving that binary stream, would de-serialize and decode it according to interpret the byte stream into an understandable format. Next, the user program would perform the target computation (User Logic), which ranges from simple tasks such as rotate the jpg file by 90 degrees if needed, to relatively complex tasks such as detecting pedestrians given the binary sensor readings from LiDAR scanners. The output would then be encoded and serialized before being passed in the form of RDD[Bytes] partitions. In the last stage, the partitions can be returned to the Spark driver through a collect operation or be stored in HDFS as binary files. With this process, we can now process and transform binary data into a user-defined format and transform the output of the Spark computation into a byte stream for collect operations or take it one step further to convert the byte stream into text or generic binary files in HDFS according to the needs and logic of applications.

### 3.2 Data Retrieval through ROSBag Cache

In this subsection, we present our design of reading from the memory of the cache data through ROSBag play function, and of caching data into memory through ROSBag record function. As shown in Figure 5, in our current design, ROSPlay takes ROSBag data as input, which is passed to ROS through BinPipeRDD. Once done with simulation, ROSRecord can persist the output through BinPipeRDD to some form of customized data format.

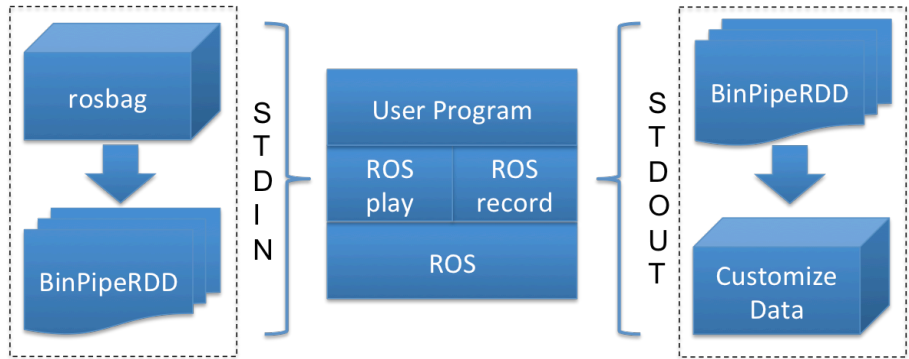

**Fig 5. Simulation workflow**

But the missing links still exist in this process, including how ROSBag play function reads the cached data from memory, and how the ROSBag record function caches the data into memory. In order to realize these functions, we add a branch logic layer for the original two-layer logical structure of Bag and ChunkedFile (see Figure 2 for more details). As shown in Figure 6, the MemoryChunkedFile class inherits from the ChunkedFile class and overrides all the methods of ChunkedFile. MemoryChunkedFile reads and writes files to the lower layer's memory, but not reads and writes data to the hard disk as the ChunkedFile class does. A major benefit of this design is that the worker can read data passed to simulators through standard input stream directly instead of reading and writing through Disk I/O. Instead of passing data through disk I/O, with this design, we can pass data through memory directly. Therefore, due to acceleration from the I/O side, this read/write model greatly reduces the time of data processions.

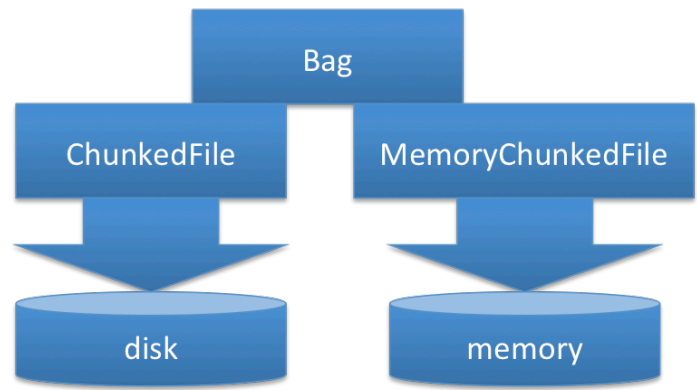

**Fig 6. MemoryChunkedFile Design**

With the addition of this logic layer, we can now deploy the simulator to every worker machine in the Spark cluster. By running different configuration files we can make each machine runs a different module. Or we can deploy the same modules and models under different conditions to run the same data to compare the differences between these models. In addition, we can also deploy the same modules and models under the conditions of running different data to compare different data. Thus, the use of distributed systems greatly enhances the performance and the flexibility of the simulation platform design.

## 4 Performance Evaluation

In this section we delve into the performance of our simulation platform. Since the platform is mainly for accelerating simulation workloads on the cloud, we study the I/O performance of our platform, as well as its scalability.

### 4.1 ROSBag Cache Performance

As shown in Figure 6, to test the performance of ROSBag cache, we compare the performance of ROS play (read) and ROS record (write) with and without using in memory cache. We perform two test cases, the Small File Test, which repeatedly read and write 1 million files with 1 KB in size, and the Large File Test, which repeatedly read and write 100 thousand files with 1 MB in size. The no cache case uses the original ChunkedFile whereas the with cache case uses the MemoryChunkedFile. We perform this test on a 12-core server machine with 65 GB of main memory. The results show that with in-memory cache, the write performance gets improved by about 3X and the read performance gets improved by 5X in the large file test, by about 10X in the small file test. This result confirms that the MemoryChunkedFile is an effective way to improve I/O performance in our simulation tests.

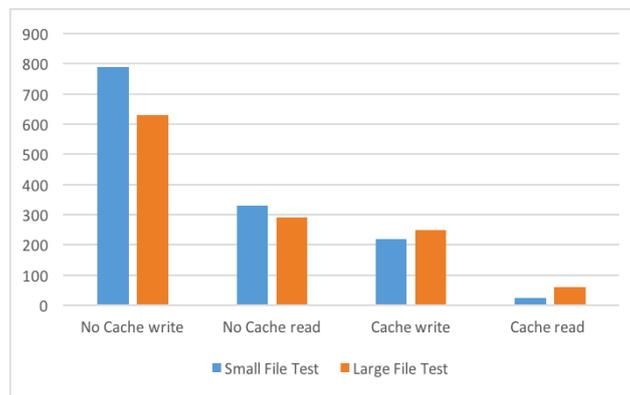

**Fig 6. System Scalability**

## 4.2 Scalability

As shown in Figure 7, we conduct a scalability evaluation of the system. With the increase of computing resources, the calculation time is also linearly reduced. The system shows a strong scalability. In an internal image recognition test set, it takes 3 hours to process images using stand-alone processing, and only 25 minutes after using eight Spark workers. We don't have the Google autonomous driving dataset, but let us extrapolate this study and apply on Google dataset. Suppose that if we use 10000 Spark workers to test large-scale image recognition simulation on Google's unmanned cars data, the entire experiment can be done in 100 hours, whereas on a single machine, this it will take more than 600,000 hours to complete.

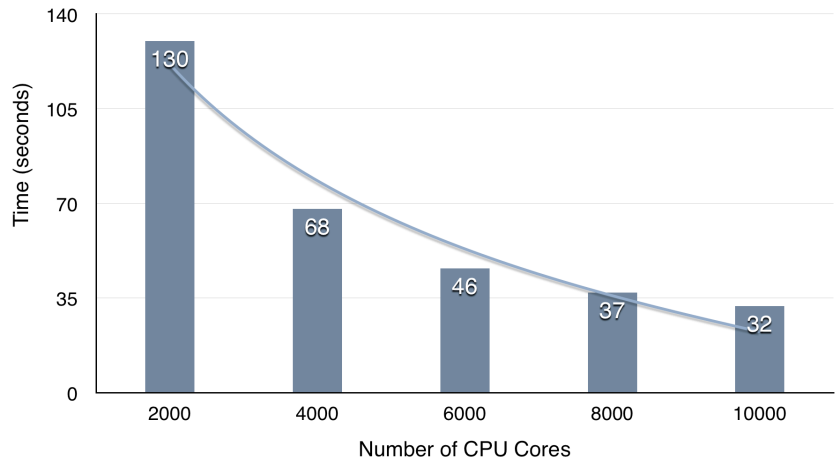

**Fig 7. System Scalability**

## 5 Related Work

In this paper, we present a general distributed simulation framework based on Spark distributed computing system and ROS. Note that our platform is general in the way that the simulator, in this case ROS, can be replaced by any other simulators. Previously there are several autonomous vehicle simulators developed. The simulators used by car makers include IPG Automotive GmbH and VEDYNA, [4] and [5], which provide numerical simulations of full car dynamics with interfaces to MATLAB/Simulink. Both of these simulators try to ease the development and integration of vehicle controllers. Similarly, the ADTF can be used to model a directed graph reflecting the data flow through a set of processing modules [6]. The communication is realized using so-called channels, which themselves are typed but which can carry arbitrary typed data in principle contrary to the approach realized in the software framework Hesperia which relies solely on typed messages instead. Additional to the aforementioned ADTF, the toolkit Virtual Test Drive is developed to manage previously recorded raw sensor data or to synthetically generate required input data to perform simulations [7]. TNO PreScan can be used to support the development of so-called pre-collision driver assistance systems [8]. Another approach is provided by a tool from IAV [9]. This tool generates synthetic raw data for arbitrary sensors. Therefore, the user models in a 2D manner the characteristics of a specific active sensor like a field of view (FOV), a maximum distance, and some error noise. Then, the software computes preprocessed sensor data which would be provided by the ECUs of a specific sensors. FastSim is an open-source lightweight simulation environment designed to facilitate motion planning algorithm development for urban autonomous driving [10], which can be used to simulate the decision algorithms of autonomous vehicles.

# 6   Conclusions

Traditionally, autonomous vehicle algorithm simulations run on single machines, which takes enormous amount of time to finish. In addition, it takes increasingly more time to perform simulations as the system becomes more complex. Therefore, to accelerate the simulation process, we utilize a distributed computing framework. In this paper we present a production distributed simulation framework based on Spark, which is used for distributed computing, and ROS, which is used for playback simulations. To enable such a distributed simulation platform, we need to seamlessly integrate Spark and ROS, as well as have Spark consuming multimedia data. We also demonstrate that the system exhibits very good scalability, such that as we provide more computing resources, the simulation time drops almost linearly. Note in this paper we only demonstrate the ROS-based playback simulator. However, the proposed simulation platform is generic, such that we can plug in any other simulator to perform distributed simulation. We believe that this platform will become a standard service of the autonomous driving cloud [11].